\documentclass[
twocolumn,
superscriptaddress,
nofootinbib,
 amsmath,amssymb,
 aps,
pra,
]{revtex4-2}

\usepackage{graphicx}% Include figure files
\usepackage{dcolumn}% Align table columns on decimal point
\usepackage{bm}% bold math
\usepackage{tabularx} % for two-column things

\usepackage{tikz}
\usepackage{float}
\usetikzlibrary{quantikz}

\newcommand{\m}{\mathcal}

\newcommand{\la}{\langle}
\newcommand{\ra}{\rangle}
\newcommand{\kb}{\ra\la}
\newcommand{\half}{\frac{1}{2}}

\newcommand{\subs}[1]{$_{#1}$}

\begin{document}

\preprint{APS/123-QED}

\title{Particle-conserving quantum circuit ansatz with applications in variational simulation of bosonic systems}

\author{Sina Bahrami}
 \affiliation{Intel Corporation, Logic Technology Development}
 \email{sina.bahrami@intel.com}

\author{Nicolas Sawaya}
 \affiliation{Intel Labs (former affiliation)}
 \affiliation{Azulene Labs Inc.; \textit{and} HPI Biosciences Inc.}
 \email{nicolas@azulenelabs.com}

\date{\today}% It is always \today, today,

\begin{abstract}
Constrained problems are frequently encountered in classical and quantum optimization. Particle conservation, in particular, is commonly imposed when studying energy spectra of chemical and solid state systems. 
Though particle number-constraining techniques have been developed for 
fermionic (\textit{e.g.} molecular electronic structure) Hamiltonians, analogous techniques are lacking for non-binary and non-fermionic problems, as in the case of bosonic systems or classical optimization problems over integer variables.
Here we introduce the binary encoded multilevel particles circuit ansatz (BEMPA)---an ansatz which preserves particle count by construction---for use in quantum variational algorithms. The key insight is to build the circuit blocks by carefully positioning a set of symmetry-preserving 2- and 3-qubit gates.
We numerically analyze the problem of finding the ground state eigenvalues---via the Variational Quantum Eigensolver (VQE) algorithm---of the Bose-Hubbard Hamiltonian. For a range of model parameters spanning from Mott insulator to superfluid phase, we demonstrate that our proposed circuit ansatz finds the ground state eigenvalues within drastically shorter runtimes compared to penalty-based strategies methods. 
Finally, we analyze the potential resource benefits of changing the qubit encoding at the end of the optimization routine. 
Our results attest to the efficacy of BEMPA for simulating bosonic problems for which particle number is preserved. 

\end{abstract}

\maketitle

% ****************
%\section{Introduction}
% ****************

{\it Introduction.} One of the most promising applications of digital quantum computers is the simulation of quantum systems \cite{mcardle2020review,childs2021theory,di2023quantum}. 
When designing quantum algorithms for microscopic physical simulations, the methods applied to fermionic \cite{abrams1997simulation,whitfield2011simulation} systems may not be readily applicable to bosonic \cite{sawaya2020dlev} ones, as the two differ fundamentally in their quantum mechanical formulations. This is largely because the two particle types have different commutation relations and allowable occupation numbers, which in turn results in different requirements for qubit encodings. 

For simulations running on early generations of quantum hardware, two classes of hybrid algorithms  stand out in particular; namely the Variational Quantum Eigensolver (VQE) and its variants \cite{peruzzo2014variational,mcclean2016njp,fedorov2022vqe}, and the Quantum Approximate Optimization Algorithm (QAOA) \cite{farhi2014qaoa}. Implementing these algorithms requires some arbitrary choices for
the qubit encoding and circuit ansatze, as well as classical optimization routine. These algorithmic components require careful design especially when only a fraction of quantum computer's available Hilbert space is valid, for example when the wavefunction is constrained to a particular number of particles.

There are multiple choices for encoding $d$-level variables or modes into qubits, perhaps most prominently the standard binary, Gray, and unary (one-hot) encodings \cite{somma2003physics,mcardle19_qvibr,sawaya2020dlev,glos2022space,sawaya2022dqir}.
Notably, the unary (or one-hot) encoding easily allows for particle conservation via only two-qubit gates \cite{sawaya2020dlev}. 
However, the unary code's main drawback is that it requires $d$ qubits for each quantum degree of freedom, with $d$ being the cardinatlity of the quantum variable, {\it e.g.} the tally of available number states for a truncated bosonic mode. Alternatively, compact codes such as standard binary or Gray require $\log_2 d$ qubits, but current methods for preserving particle count requires highly non-local operators across multiple qubits, which in turn leads to significant circuit depths.
Ideally, one would like to \textit{both} use a memory-efficient compact qubit representation ({\it e.g.} Gray or standard binary) and maintain the ability to efficiently constrain particle count.

In this letter, we introduce the binary encoded multi-level particles ansatz (BEMPA). Here ``multi-level'' refers to variables with cardinality $d$ larger than 2 as in the case of bosonic modes. %, or colors in a classical k-coloring graph problem. 
When applied to certain constrained systems ({\it e.g} microscopic solid states models obeying particle conservation or classical optimization problems with similar constraints \cite{papadimitriou1998book,meyer2023exploiting}), we expect this novel ansatz to substantially reduce the convergence time for VQE and QAOA. By design, all one-, two-, and three-qubit operations in BEMPA obey particle conservation which eliminate the need to use penalty-based regularizations that can be computationally inefficient.
Hence BEMPA allows for efficient particle conservation while simultaneously allowing for the use of compact encodings (binary and Gray) that require $\log_2 d$ qubits per mode.

The past literature is replete with notable works on simulating bosonic models and bosonic field theories \cite{jordan2011scatter,jordan2012science_qft,somma2016_1dhamsim_qhos,macridin2018pra,macridin2018prl,klco2019scalarfields,sawaya2020dlev,sawaya2020connec,macridin2021bosonic,liu2021towards,tong2021provably}.  
Perhaps most relevant to this letter is the work of Liu {\it et al.} \cite{liu2021towards} where near-term quantum algorithms such as VQE and Quantum Imaginary Time Evolution (QITE) were considered and a bosonic unitary coupled cluster technique was introduced. In a related direction, there has been some published work regarding quantum simulation of phononic and vibrational degrees of freedom  \cite{mcardle19_qvibr,ollitrault20_reiher_qvibr,sparrow18_vibrdyn,veis2016quantum,sawaya2021pra,sawaya2023notrap} which uses operators with the same bosonic commutation relations (however, most such problems do not require particle conservation). Other notable works in this space include Hamiltonian simulations on qu\textit{d}its \cite{kurkcuoglu2021_phi4_qudit,deller2022quditqaoa} as well as VQE methods for preserving arbitrary symmetries of a given Hamiltonian class \cite{gard2020symm,seki20_symmadapted,barron21_symmnoise,lyu2022symmetry,lacroix2023symmetry,picozzi2023symmetry}. Also notable is the literature on symmetry-preserving ansatze for a range of other applications \cite{skolik2022equivariant,anselmetti2021qnumbfermion,sawaya2022dqir,meyer2023exploiting}.

%
%\subsection{Hamiltonian}
%
{\it Hamiltonian model.} As an example application, we analyze models of self-interacting ultracold bosonic atoms at zero temperature in 1D optical lattices that can be described using the Bose-Hubbard (BH) Hamiltonian \cite{PhysRevB.40.546,PhysRevLett.81.3108}
\begin{equation} \label{BHham}
\hat{H}_{\rm bh} = - \mu \sum_{i=1}^{N} \hat{n}_i - \omega_{t} \sum_{i \neq j}^{N} \ [\hat{a}_i ^\dagger \hat{a}_j + \hat{a}_j ^\dagger \hat{a}_i] + \omega_{\rm int} \sum_{i=1}^{N} \hat{n}_i(\hat{n}_i-1).
\end{equation}
Here $\hat{n}_{i} := \hat{a}^{\dagger} _i \hat{a}_i$ is the number operator at the $i$th site with the creation and annihilation operators $\hat{a}^{\dagger} _i$ and $\hat{a}_i$ satisfying the standard commutation relations $[\hat{a}_i, \hat{a}^{\dagger} _j] = \delta_{ij}$ and $N$ is the total number of cavities or modes (we set $\hbar = 1$ hereon). The parameter $\mu$ is the chemical potential, $\omega_t$ is the tunneling frequency between adjacent cavities which is multiplied by the so-called ``hopping'' term, and $\omega_{\rm int}$ controls the strength of self-interactions at any given lattice site. A feature of Bose-Hubbard models is the existence of two dynamical phases; when self-interactions dominate over tunneling, {\it i.e.} $\omega_{\rm int}/\mu \gg \omega_t/\mu$, the bosonic particles populate insulating Mott states, while they form a superfluid state when the reverse is true. In the mean field approximation, it can be shown that a phase transition occurs when $\omega_{\rm int}/\omega_t \sim n  (1+\sqrt{1+1/n})^2$ where $n$ is the average number of quanta per site (we have omitted the lattice coordination number which is $\mathcal{O}(1)$ for our models). Our numerical investigations will examine a range of Hamiltonian parameters corresponding to both quantum phases. We use this model as a prototypical example, but we note that BEMPA is applicable to any U(1)-preserving Hamiltonian class, potentially including classical optimization problems for which the variable sum is constrained ({\it e.g.} some versions of the bin packing problem \cite{papadimitriou1998book}).

\begin{figure}
    \centering
    \includegraphics[width=.95\linewidth]{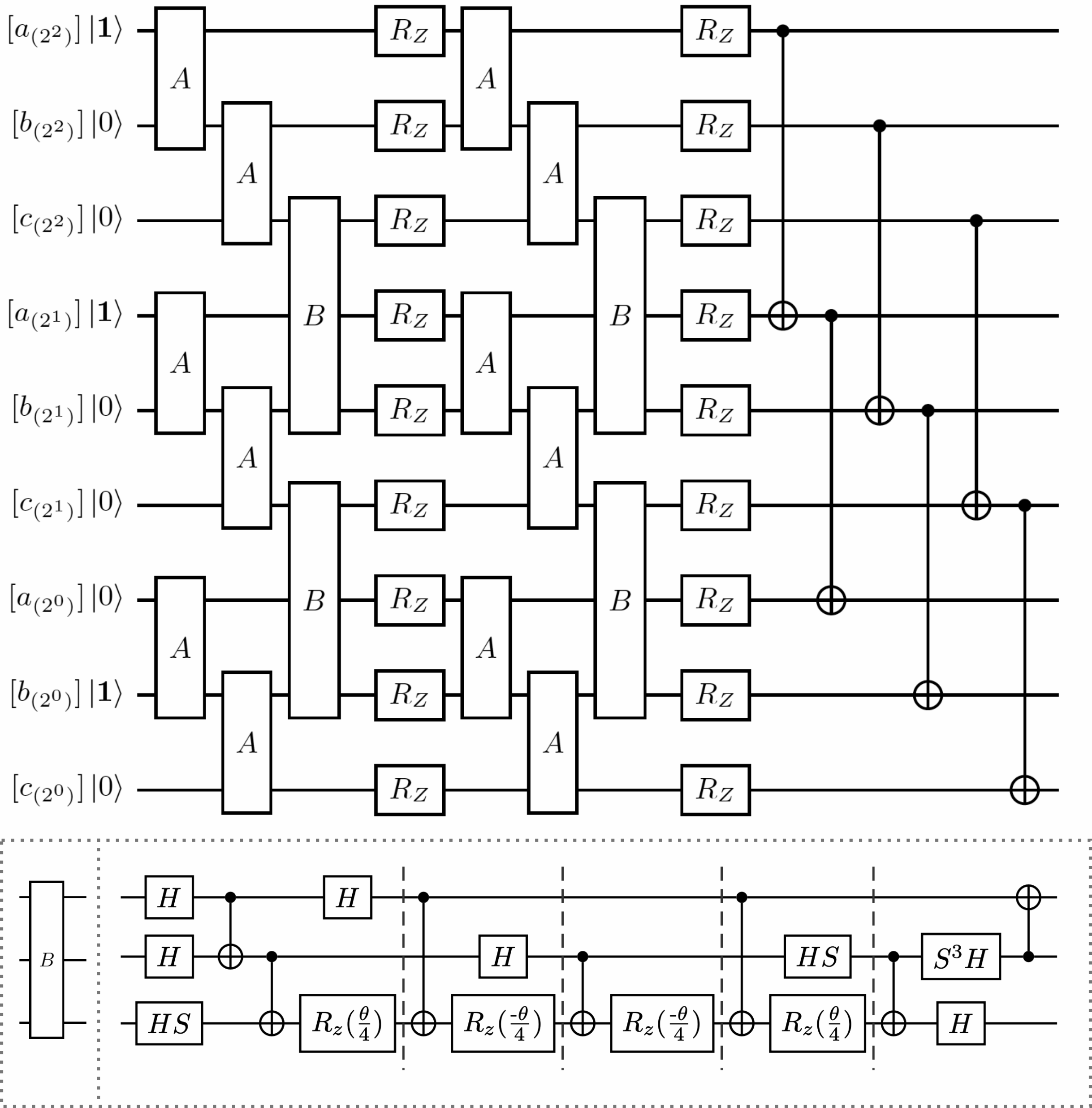}
    \caption{Top figure is a sample illustration of the binary encoded multilevel particles ansatz (BEMPA). In this example, particles $a, b,$ and $c$ are initialized with 6 (110$_2$), 1 (001$_2$), and 0  (000$_2$) quanta respectively. By design, the total number of quanta is preserved by this circuit ansatz (note the careful ordering of the qubits and placement of the gates). The final sequence of CNOT gates is an optional procedure that converts the standard binary to the Gray code. The low-depth $\hat{B}$ gate decomposition derived in this work is shown in the bottom figure. 
    See reference \cite{gard2020symm} for more details on the $\hat{A}$ gate.
    }
    \label{fig:bempa}
\end{figure}

% ****************
%\textit{Methods.} 
% ****************

\begin{figure*}[!t]
    \centering
    \includegraphics[width=1\textwidth]{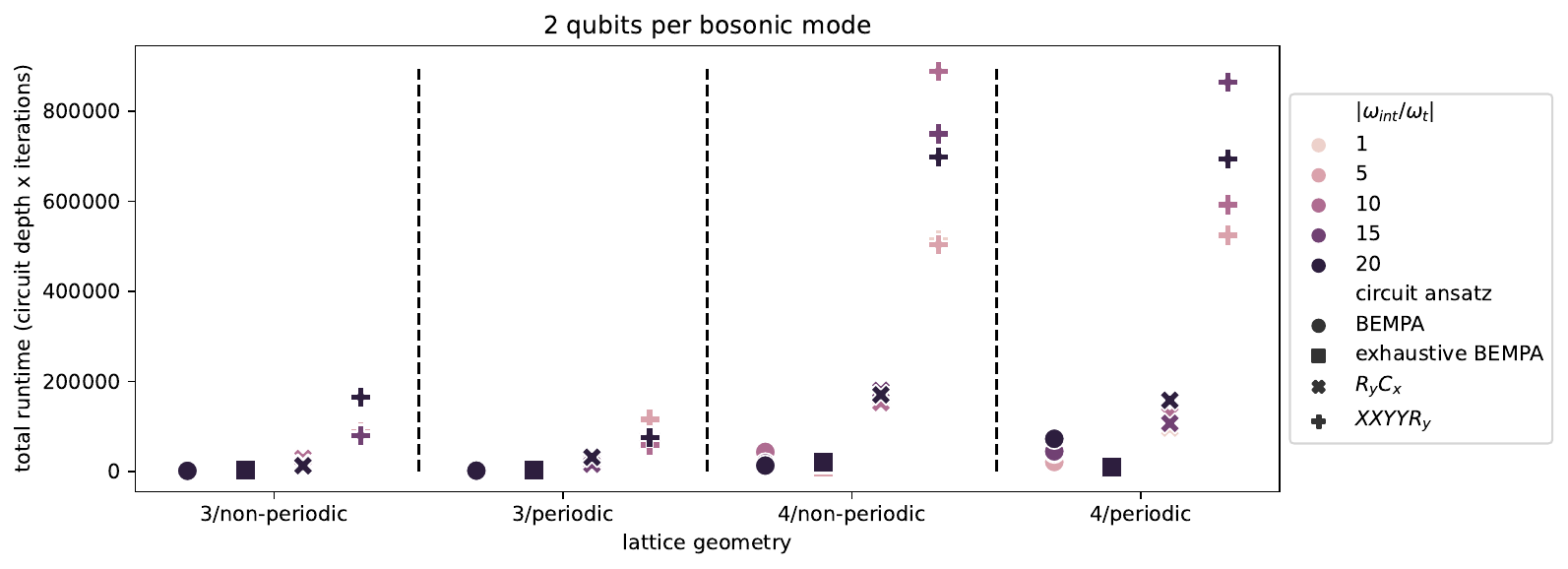}
    \caption{Total runtimes 
    are compared for both versions of BEMPA and two other circuit ansatze used in the penalty-based methods for a range of $\omega_{\rm int}/\omega_{\rm t}$ that covers the Mott insulator to superfluid phases. The simulated models include three and four bosonic modes trapped in a 1D lattice both with and without periodic boundary conditions. For each model, VQE is used to estimate the minimum eigenvalue of quantum states with the occupation number of $(n \times k) + 1$, where $n$ is the number of modes and $k$ is the number of qubits per mode. The penalty value $\eta$ is fixed to $10$ for all simulations. 
    }
    \label{fig:vary}
\end{figure*}

\textit{Methods.}  When implementing VQE for a problem with symmetry constraints, the default strategy is to introduce penalty terms \cite{mcclean2016njp,ryabinkin2018constrained}. One such method involves replacing the original Hamiltonian $\hat{H}$ with an effective Hamiltonian
\begin{equation}\label{eq:penalty}
\hat{H}_{\rm eff} = \hat{H} + \hat{P}
\end{equation}
where $\hat{P}$ is an operator that penalizes deviations from the allowable subspace. To conserve particle count, one may use $\hat{P}=\eta \big( \hat{N}_{\rm tot} - \m N \big)^2 = \eta \big (\sum_i \hat{n}_i - \m N \big)^2 $, where 
$\hat{N}_{\rm tot}$ is the total particle number operator, $\m N$ is the number of particles for the desired eigenstate, and $\eta$ is an arbitrary real constant. 

A main drawback of penalty-based methods is that they leave the entire Hilbert space accessible. In other words, the optimizer is not limited to searching within a subspace of states that observe the symmetry constraints of the problem. 
The purpose of this letter is to introduce circuit ansatze that incorporate particle conservation at the {\it kinematical} level, 
leading to circuits with fewer gates, fewer parameters, and a significantly smaller portion of the Hilbert space for the optimizer to traverse. 
The BEMPA ansatz assumes the use of the standard binary encoding; the number of qubits depends on the number of energy levels to which the mode has been truncated. For conceptual convenience and for reducing SWAP gate requirements, we organize the qubits into ``significant figure blocks" (SFBs) such that for the same significant figure $k$ the qubits corresponding to $2^{k}$ are clustered together (see Figure \ref{fig:bempa}). Then, within each block, a gate  $\hat{A}$ may be implemented with sparsity pattern (see \cite{gard2020symm} for more details)
\begin{equation}\label{eq:A}
\hat{A} = \begin{pmatrix}
1 & 0 & 0 & 0 \\
0 & * & * & 0 \\
0 & * & * & 0 \\
0 & 0 & 0 & 1 \\
\end{pmatrix}.
\end{equation}
Because the number of particles is conserved within each block, this ensures that particle count is conserved for the entire system.

However, a circuit composed entirely of $\hat{A}$ gates can access only a fraction of states with a fixed particle count.
To understand this point, consider $|\psi_a\ra = |1\ra_a |1\ra_b |0\ra_c \rightarrow |001_2\ra_a |001_2\ra_b |000_2\ra_c$ and $|\psi_b\ra = |0\ra_a |0\ra_b |2\ra_c \rightarrow |000_2\ra_a |000_2\ra_b |010_2\ra_c$, where each state is encoded as an 8-level system using the standard binary encoding denoted by subscript 2. Clearly, projecting $|\psi_a\ra$ to $|\psi_b\ra$ cannot be done if one were to rely on $\hat A$ gates alone, despite both states having the same total number of quanta $\m N = 2$.
Hence we require gates that allow transfer of amplitude between different significant figure blocks. To this end, we introduce a class of gates $\hat{B}$ with sparsity pattern

\begin{equation}\label{eq:B}
\hat{B} = \begin{pmatrix}
1 & 0 & 0 & 0 & 0 & 0 & 0 & 0 \\
0 & * & 0 & 0 & 0 & 0 & * & 0 \\
0 & 0 & 1 & 0 & 0 & 0 & 0 & 0 \\
0 & 0 & 0 & 1 & 0 & 0 & 0 & 0 \\
0 & 0 & 0 & 0 & 1 & 0 & 0 & 0 \\
0 & 0 & 0 & 0 & 0 & 1 & 0 & 0 \\
0 & * & 0 & 0 & 0 & 0 & * & 0 \\
0 & 0 & 0 & 0 & 0 & 0 & 0 & 1 \\
\end{pmatrix}.
\end{equation}
This unitary operator transfers amplitude between two $2^{k-1}$ bits and a single $2^k$ bit. For example a $\hat{B}$ gate in Figure \ref{fig:bempa} mixes $| 0\ra_b^{(k-1)}|0\ra_a^{(k-1)}|1\ra_c^{(k)} $ and $| 1\ra_b^{(k-1)}|1\ra_a^{(k-1)}|0\ra_c^{(k)}$, leaving other states untouched, where we have added superscripts to denote each qubit's SFB. The idea is to always position these gates such that if the first two qubits correspond to significant figure $k$, then the second qubit corresponds to $k+1$. In Figure \ref{fig:bempa} for example, the first $\hat{B}$ gate is placed such that $k=1$, spanning the $2^2$ and $2^1$ blocks. 
Because $\hat B$ allows for amplitude transfer between SFBs, we postulate that an ansatz composed of both $\hat{A}$ and $\hat{B}$ is sufficient to reach any state with the same particle count as the initial state.

It is useful to derive the Pauli representations of Hermitian generators that produce unitaries of type $\hat A$ and $\hat B$. In fact, explicit representations of such generators may be strictly required in some near-term quantum algorithms, for example in the more advanced ADAPT-VQE \cite{grimsley2019adaptive} and in some implementations of QITE \cite{motta2020determining,sun2021quantum}. 
It can be shown that the generators for $\hat A$ and  $\hat B$ are $\hat G_{A} = i| 01 \kb 10 | - i| 10 \kb 01 |$ and $\hat G_{B} = i| 001 \kb 110 | -i| 110 \kb 001 |$ respectively. By substituting $|0\kb1|=\half (\hat X+i\hat Y)$ and $|1\kb0|=\half (\hat X-i\hat Y)$ for each qubit, we used \texttt{mat2qubit} \cite{mat2qubit} to derive the Pauli representations

\begin{equation}\label{eq:GA}
\hat G_{A} \rightarrow \half \left (  \hat X_0 \hat Y_1 - \hat Y_0 \hat X_1  \right ),
\end{equation}
\begin{equation}\label{eq:GB}
\hat G_{B} \rightarrow \frac{1}{4} \left ( \hat X_0 \hat X_1 \hat Y_2 - \hat X_0 \hat Y_1 \hat X_2 - \hat Y_0 \hat X_1 \hat X_2 - \hat Y_0 \hat Y_1 \hat Y_2  \right ).
\end{equation}

The finite transformations given in Eqs. \eqref{eq:A} and \eqref{eq:B} may now be expressed as $\hat A(\theta) = \exp(-i \theta \hat G_A)$ and $\hat B(\alpha) = \exp(-i \alpha \hat G_B)$. Although the unitaries 
that result from these generators are strictly real, an imaginary component can be introduced either by using a different generator or by incorporating $Z$-Rotation gates. It is important to note that these particle-conserving operators are exactly implementable with first-order Suzuki-Trotter formulae \cite{suzuki1985decomposition} as all terms commute. 

However, utilizing standard ``CNOT ladder'' \cite{mikeandike2010} constructs requires substantial circuit depths. In this work (see Appendix \ref{sec:decomp}) we instead derive a shorter-depth circuit decomposition of $\hat B$, based on Pauli tableau (or Pauli frame) representations \cite{schmitz2021popr,aaronson2004improved,maslov2018shorter}. The bottom panel of Figure \ref{fig:bempa} shows our depth-13 decomposition, significantly shorter than the depth-25 decomposion of the CNOT ladder construction. We used CNOTs and arbitrary one-qubit gates as our gate set.

\begin{figure}
    \centering
    \includegraphics[width=0.5\textwidth]{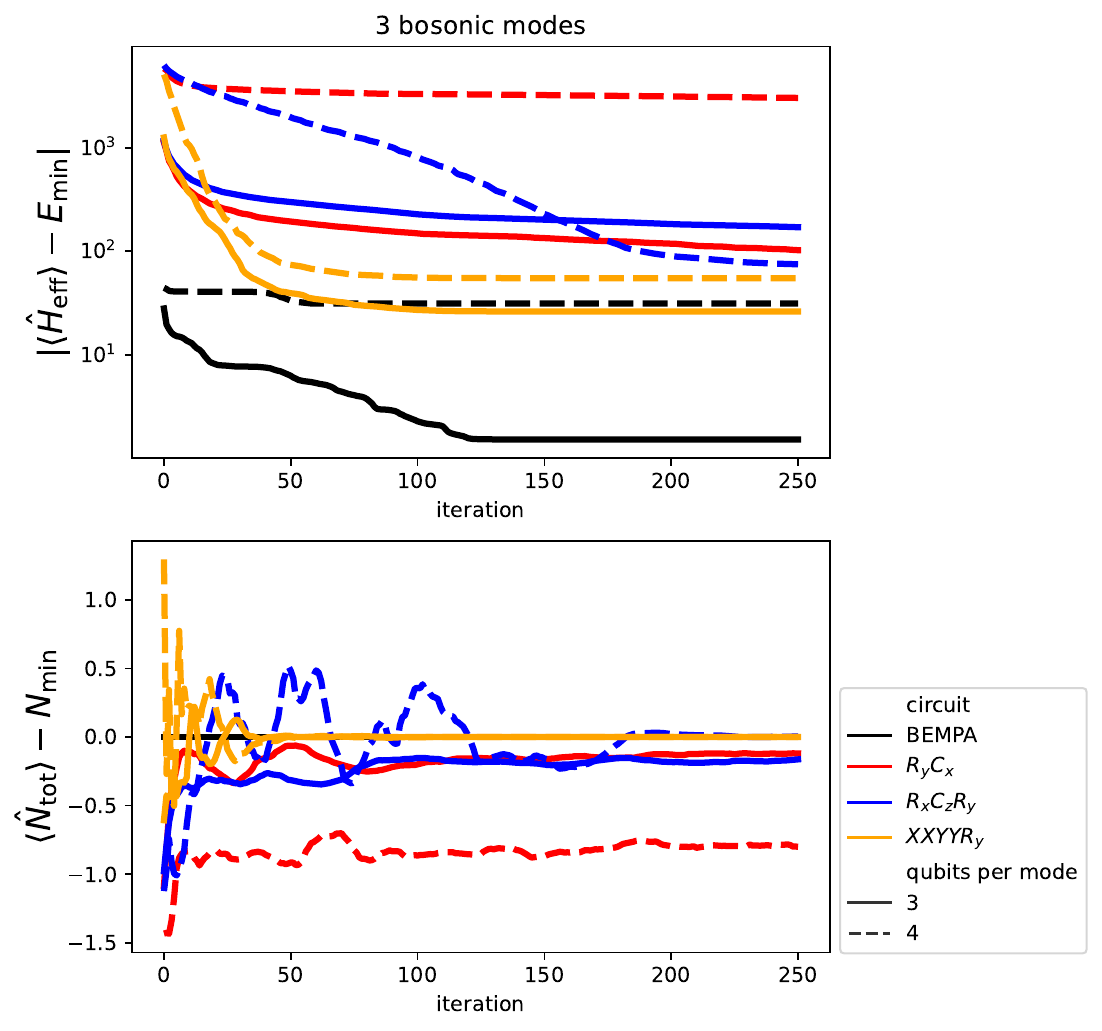}
    \caption{Absolute error with respect to the effective Hamiltonian as well as difference between the total particle count and the minimum state's particle count are plotted versus optimization step for the circuits under study. The models considered here are 1D arrays of 3 bosonic modes with $d=3$ and $d=4$, occupation numbers of $3 \times \log_2{d}$, and dynamics governed by the BH Hamiltonian with $\mu = |\omega_{\rm int}| = |\omega_t|=1$. The circuits are truncated at the maximum depths of 100 and 70 for $d=3$ and $d=4$ models respectively. As before, the penalty parameter $\eta$ is set to 10 for all simulations plotted here. Expectedly, the bottom plot indicates triviality of the penalty term for BEMPA while for penalty-based optimizations it shows the optimizer's freedom in traversing eigenstates with different total particle counts.}
    \label{fig:iterationplot}
\end{figure}

Finally, we discuss two reasons for why converting encoding from standard binary to Gray at the end of the BEMPA ansatz, by appending the depth-$O(\log_2 d)$ circuit \cite{sawaya2020dlev} of CNOTs shown in Figure \ref{fig:bempa}, can prove advantageous in some scenarios. 
First, reproducing $\la \hat{H} \ra$ in practice requires numerous repetitions of quantum circuit ({\it e.g.} $10^9$ repetitions even for relatively small systems \cite{gonthier2020meas}).
Converting to the Gray code may often lead to fewer circuit repetitions, mostly because the Gray code produces Hamiltonians that are more local, by which we mean the terms have lower Pauli weight. Second, for many applications, one may want to run dynamics after preparing the state with BEMPA, and the required time-dependent exponential $\exp(-i \hat H \tau)$ is expected to be more efficient in Gray than in standard binary \cite{sawaya2020dlev}. In order to investigate the potential benefits of this encoding conversion when simulating the BH model with BEMPA, in Appendix \ref{sec:measreduc} we present encoding-dependent results for Pauli term counts, Pauli weight, and shot counts.

% ****************
\textit{Numerical simulations.} 
% ****************
We simulate VQE for three and four bosonic modes in one dimensional lattices with dynamics described by the BH Hamiltonian given in Eq. \eqref{BHham}. Our choices of circuit ansatze used for the penalty-based simulations are shown in Appendix \ref{sec:ansatze}.  
In addition to the BEMPA circuit with the $\hat{B}$ gate pattern shown in Fig. \ref{fig:bempa}, another variation is considered in which every $\hat{A}$ gate is immediately followed by $\hat{B}$ gates that entangle the level-$k$ qubits coupled by that instance of $\hat{A}$ gate and any other level-$k+1$ qubit. We call this variation the exhaustive BEMPA. 
Logistically, qubit Hamiltonians are taken from \texttt{HamLib} \cite{sawaya2023hamlib} or prepared using \texttt{mat2qubit} \cite{mat2qubit} and the simulations are performed using an in-house code based on \texttt{Scipy} \cite{scipy}. The standard BFGS algorithm \cite{jorge2006book} is used with the step-size and gradient norm tolerance parameters tuned to $10^{-8}$ and $10^{-10}$ respectively. This tuning is based on our convergence criterion, which is taken to be within $10^{-8}$ of the desired eigenvalue; it was set while minimizing the required BEMPA circuit depth for a sample BH model with two 4-level bosonic modes.

The numerical results shown in Fig. \ref{fig:vary} show the BEMPA circuits substantially outperforming the selected non-particle-conserving circuits over a range of values in the BH parameter space, considering three and four 4-level bosonic modes arranged in a 1D lattice with and without periodic conditions. We compare BEMPA to circuits composed of what we call $R_y C_x$, $R_x C_z R_y$, and $XXYYR_y$ blocks (see Appendix \ref{sec:ansatze}). We chose the former two penalty-based ansatzae because they are very similar to ubiquitously used ansatzae \cite{nakaji2021expressibility, cerezo2021cost, wierichs2020avoiding, bergholm2018pennylane}. The latter penalty-based ansatz, $XXYYR_y$, is an attempt to provide a deliberately contrived ansatz that is more competitive with BEMPA, in order to see whether BEMPA remains the most performant. Because $e^{-i \theta (XX + YY)}$ preserves particle count, the idea is that this ansatz will more easily remains in the correct particle count manifold.

In each case, the simulations are run with $|\omega_{\rm int}/\omega_t|$ $\in \{1,5,10,15,20\}$, which subsumes a range of parameters extending from the Mott insulator to superfluid phase \cite{Freericks_1994}. Evidently, the gap in total runtime between BEMPA and other ansatze becomes more pronounced as the complexity of model increases. 

Perhaps it is more insightful to compare the optimizer error as a function of iterations. This is illustrated for a few sample simulations in Fig. \ref{fig:iterationplot}. Here we use larger BH models consisting of three 8-level/16-level bosonic modes in a 1D lattice without periodic boundary conditions with the Hamiltonian parameters set to unity. For a more even comparison, we also cap the circuit depths at 70 and 100 for the 16-level and 8-level models respectively. As shown, the BEMPA circuits reduce error more substantially and within fewer iteration steps than the other circuits, a result that appears to hold irrespective of the value of the penalty parameter (not plotted). Finally, as mentioned, in Appendix \ref{sec:measreduc}, we numerically demonstrate the potential benefit (in terms of reducing shot counts) of converting encodings from standard binary to Gray at the end of the circuit.

% ****************
\textit{Conclusion.} 
% ****************

In this letter, we introduced a particle conserving circuit ansatz that will likely find suitable applications in simulating models with bosonic Hamiltonians as well as some problems in classical optimization. The primary benefit of the BEMPA circuit ansatz is that it dramatically reduces the optimization search space by restricting to states with a fixed number of particles, which in turn brings about a significantly shorter optimization runtime compared to the better known penalty-based methods as demonstrated in the context of VQE.
Indeed, the numerical results presented here show that runtime can shrink by orders of magnitude even for relatively small systems of fewer than 10 qubits.
As such, we expect this circuit ansatz to be used routinely for investigating bosonic Hamiltonians on near-term quantum hardware.

% ****************
\textit{Acknowledgements.} 
% ****************
We are grateful to Albert Schmitz for helpful discussions.

\appendix

%%%%%%%%%%%%%%%
\section{Derivation of $\hat{B}(\theta)$ decomposition}\label{sec:decomp}
%%%%%%%%%%%%%%%

In practice, it is important to implement circuits for $\hat A(\theta)$ and $\hat B(\theta)$ that are as shallow as possible. 
A depth-5 circuit for $\hat{A}(\theta)$ has been introduced in the past \cite{gard2020symm}. (Note that throughout this work we have used CNOT and arbitrary one-qubit gates as our gate set.) To arrive at a similarly optimal circuit for $\hat{B}(\theta)$, we use 
methods based on the Pauli tableau (or Pauli frame) representations \cite{schmitz2021popr,aaronson2004improved,maslov2018shorter}. The Pauli frame is given by
\begin{equation}\label{eq:pframe}
\begin{pmatrix}
 p_0 & \tilde p_0 \\
 p_1 & \tilde p_1 \\
 p_2 & \tilde p_2 \\
\end{pmatrix}
\end{equation}
where a typical initial frame (when the system is initialized as $|0\ra^{\otimes n}$) is defined by $p_i \sim Z_i$ and $\tilde p_i \sim X_i$. Quantum gates CNOT, Hadamard (H), and phase (S) change the frame according to
\begin{equation}\label{eq:framexform}
\begin{split}
 p_j,\tilde p_i &\xrightarrow{CNOT_{ij}} p_j p_i,\tilde p_i \tilde p_j \\
 % \tilde p_i \xrightarrow{CNOT_{ij}} \tilde p_i \tilde p_j \\
 p_i &\xleftrightarrow{H_i} \tilde p_i \\
 \tilde p_i &\xrightarrow{S_i} p_i \tilde p_i
\end{split}.
\end{equation}

Below we show how these frame transformations lead to the depth-13 $\hat{B}$ gate decomposition shown in Figure \ref{fig:bempa}. As mentioned in the main text, this is a notable improvement over the 25-depth decomposition that would have resulted from a naive product formula implementation.

We begin in the default frame
\begin{equation}
\begin{pmatrix}
Z_0 & X_0 \\
Z_1 & X_1 \\
Z_2 & X_2
\end{pmatrix}.
\end{equation}

Implementing ordered transformations by applying H\subs{0}, H\subs{1}, S\subs{2}, H\subs{2}, CNOT\subs{01}, and CNOT\subs{12} gates leads to the frame

\begin{equation}
\begin{pmatrix}
X_0         & Z_0 Z_1 \\
X_0 X_1     & Z_1 Z_2 \\
X_0 X_1 Y_2 & Z_2
\end{pmatrix},
\end{equation}
after which a transformation by $Rz_2(+\theta)$ gate is performed to implement the $XXY$ rotation.

Next, applying H\subs{0} followed by CNOT\subs{02} leads to
\begin{equation}
\begin{pmatrix}
Z_0 Z_1     & X_0 Z_2 \\
X_0 X_1     & Z_1 Z_2 \\
Y_0 Y_1 Y_2 & Z_2
\end{pmatrix},
\end{equation}
after which a transformation by $Rz_2(-\theta)$ gate (note the negative sign) is performed to implement the $XXY$ rotation.

Then, applying H\subs{1} followed by CNOT\subs{12} leads to
\begin{equation}
\begin{pmatrix}
Z_0 Z_1 & X_0 Z_2 \\
Z_1 Z_2 & X_0 X_1 Z_2 \\
Y_0 X_1 X_2 & Z_2 
\end{pmatrix},
\end{equation}
where similarly a transformation by $Rz_2(-\theta)$ gate (note the negative sign) is done to implement the $YXX$ rotation.

For the fourth frame, CNOT\subs{02} leads to
\begin{equation}
\begin{pmatrix}
Z_0 Z_1 & X_0 \\
Z_1 Z_2 & X_0 X_1 Z_2 \\
-X_0 Y_1 X_2 & Z_2 
\end{pmatrix},
\end{equation}
where now a transformation by an $Rz_2(+\theta)$ gate (note the negative sign in the Pauli frame representation) is performed in the frame to apply the $XYX$ rotation.

After all these transformations, we must circle back to the original default Pauli frame, which can be done with the gate sequence S\subs{1}, H\subs{1}, CNOT\subs{12}, H\subs{2}, H\subs{1}, S\subs{1}$^3$, CNOT\subs{10}. This completes the proof.

%%%%%%%%%%%%%%%
\section{Encoding change and shot reduction}\label{sec:measreduc}
%%%%%%%%%%%%%%%

\begin{figure*}[!ht]
    \centering
    \includegraphics[width=.8\linewidth]{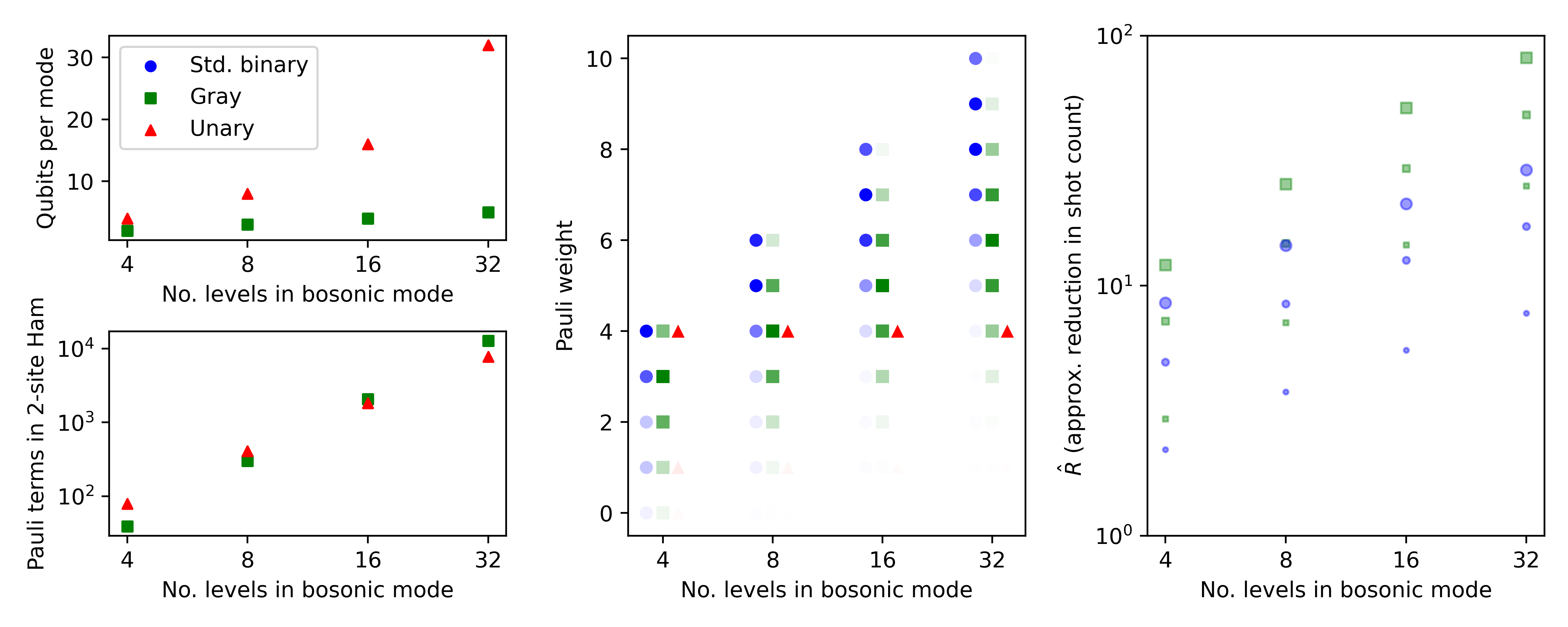}
    \label{fig:meas}
\end{figure*}

In this section we consider the potential benefits of changing the qubit encoding at the end of the BEMPA ansatz. Our primary purpose is to study the difference between Gray and standard binary. This is because, while the BEMPA ansatz requires standard binary, a very short-depth circuit allows for conversion to Gray at the end of the ansatz (this is the short CNOT pattern at the end of the ansatz in Fig. \ref{fig:bempa}). Considering the ease of this conversion, it is worth investigating whether it is beneficial. We discuss two possible benefits: fewer shot counts when estimating the energy, and more efficient dynamics.

Peripherally, we also consider the qubit-hungry unary (also called one-hot) encoding because several studies have used it in the past \cite{somma2003physics,sawaya2020dlev,ollitrault20_reiher_qvibr}. The left-hand panels of Fig. \ref{fig:meas} shows that asymptotically unary produces fewer Pauli terms in the Hamiltonian, but more qubits. We note that although unary often allows for shorter-depth circuits when all-to-all qubit connectivity is assumed, there are two major drawbacks: first, sparse hardware connectivity typically leads to deep SWAP networks when using unary \cite{sawaya2020connec}, and second, unary requires far more qubits. Hence unary will often not be a viable choice of encoding.

The center panel in Fig. \ref{fig:meas} shows the Pauli weight distribution across all terms in a BH Hamiltonian restricted to two modes. The most important trend is that the Gray code on average has a lower Pauli weight than standard binary, which tends to imply that Hamiltonian simulation may be performed in shorter circuit depth \cite{childs2021theory}. Hence, if one is using BEMPA to prepare a particular state before running dynamics, it is likely that converting to Gray would be beneficial just before beginning the dynamics simulation.

Finally, we consider the reduction in the number of shot counts required to estimate $\la \hat{H} \ra$. Many methods have been proposed for reducing the number of shot counts \cite{verteletskyi2020qwc,gonthier2020meas,crawford2021sortedins,huang2020shadow,bansingh2022fidelity,yen2021cartan,yen2020measuring,gokhale2020measN3,zhao2020measlove,dalfavero2023kqwc}; here we use the qubit-wise commuting scheme \cite{verteletskyi2020qwc} and implement the sorted insertion algorithm to calculate $\hat R$, defined in Crawford \textit{et al.} \cite{crawford2021sortedins} and shown below. $\hat R$ approximates the reduction in shot counts with respect to $N_{\text{ungrouped}}$, defined as 

\begin{equation}
N_{\text{ungrouped}} = \frac{1}{\epsilon^2} \left ( \sum_i^M a_i \sqrt{\text{Var}[P_i]} \right ),
\end{equation}
where the qubit Hamiltonian is $\hat{H} = a_i \hat{P}_i$ and $\epsilon$ is the desired error. $N_{\text{ungrouped}}$ is the number of shots required if one were to evaluate each Pauli term $\hat{P}_i$ in the Hamiltonian separately. The strategy of partitioning commuting Pauli terms together into $k$ partitions each with $m_k$ terms, such that all terms in a partition can be evaluated at once, has been shown to drastically reduce shot counts \cite{verteletskyi2020qwc,crawford2021sortedins,yen2021cartan,yen2020measuring,gokhale2020measN3,dalfavero2023kqwc}. The $\hat R$ measure, defined as
\begin{equation} \label{eqn:rhat}
    \hat{R} := \left[ 
     \frac{\sum_{k = 1} \sum_{l = 1}^{m_k} \left| a_{kl} \right|}
     {\sum_{k = 1} \sqrt{ \sum_{l = 1}^{m_k} \left| a_{kl} \right| ^ 2}}
    \right] ^ 2 ,
\end{equation}
approximates the multiplicative reduction in total shot counts for a given partitioning strategy.

Note that $N_{\text{ungrouped}}$ is equal in the standard binary and Gray encodings, so directly comparing $\hat R$ between these two encodings is a valid comparison. We omit unary from the right-hand panel because $N_{\text{ungrouped}}$ differs in the unary case, hence its value for $\hat R$ is not directly comparable to the others. (Notably, $N_{\text{ungrouped}}$ is smaller for unary than the other two encodings, and a circuit for converting from standard binary to unary is known as well \cite{sawaya2020dlev}. However, the large number of required qubits is still likely to prevent this from being a beneficial strategy.)

The right-hand panel of Fig. \ref{fig:meas} shows that, for qubit-wise commuting with sorted insertion, the conversion from standard binary to Gray leads to substantial reductions in shot counts, for a two-qubit model, a 1D periodic lattice, and a 2D periodic square lattice. These results highlight the potential benefit of converting to the Gray code at the end of the BEMPA ansatz.

%%%%%%%%%%%%%%%
\section{ansatze for penalty-based approach}\label{sec:ansatze}
%%%%%%%%%%%%%%%

\begin{figure*}\label{fig:penaltyansatze}
    \includegraphics[height=0.28\textheight,width=.3\textwidth]{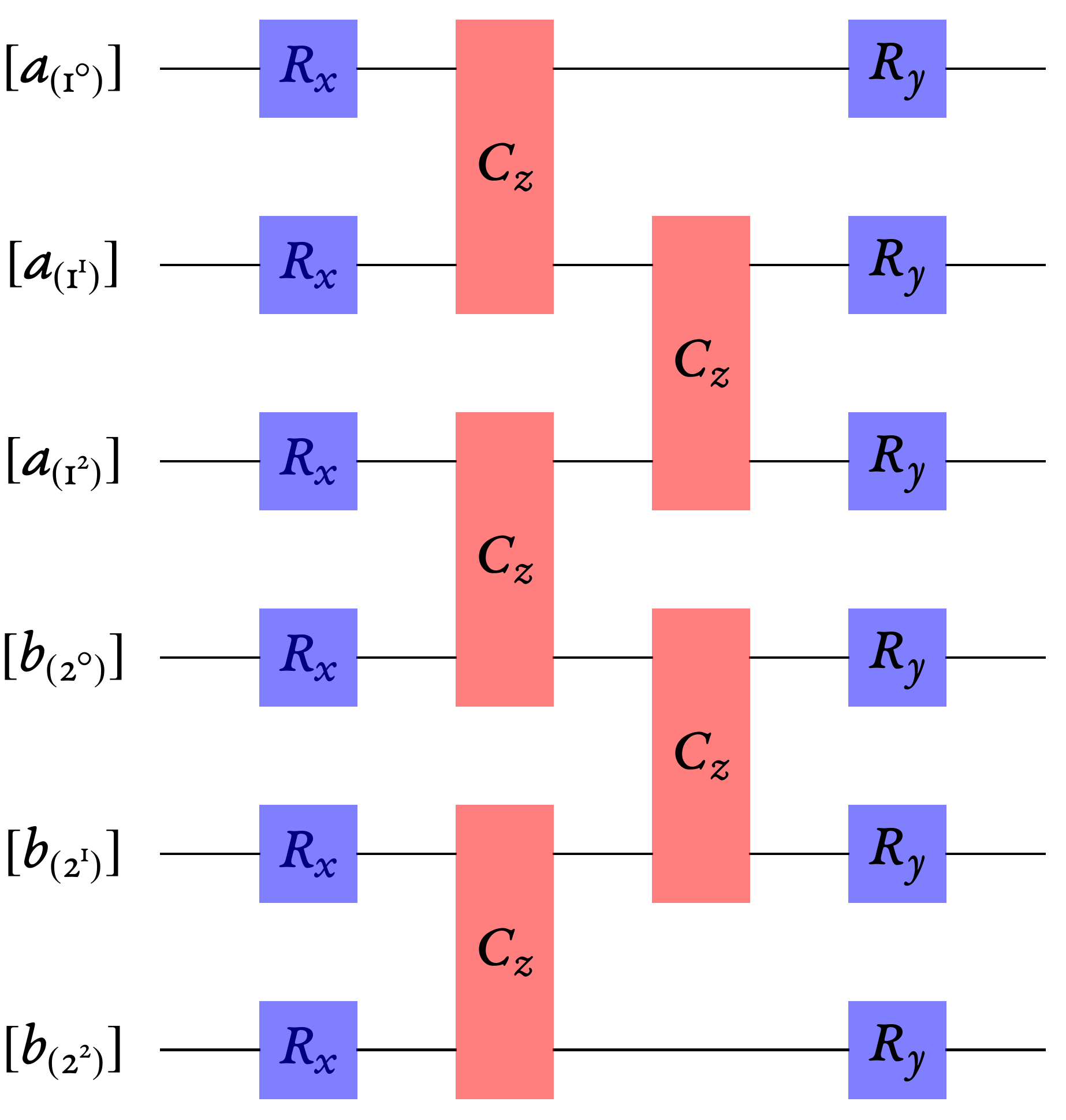}\hfill
    \includegraphics[height=0.28\textheight,width=.3\textwidth]{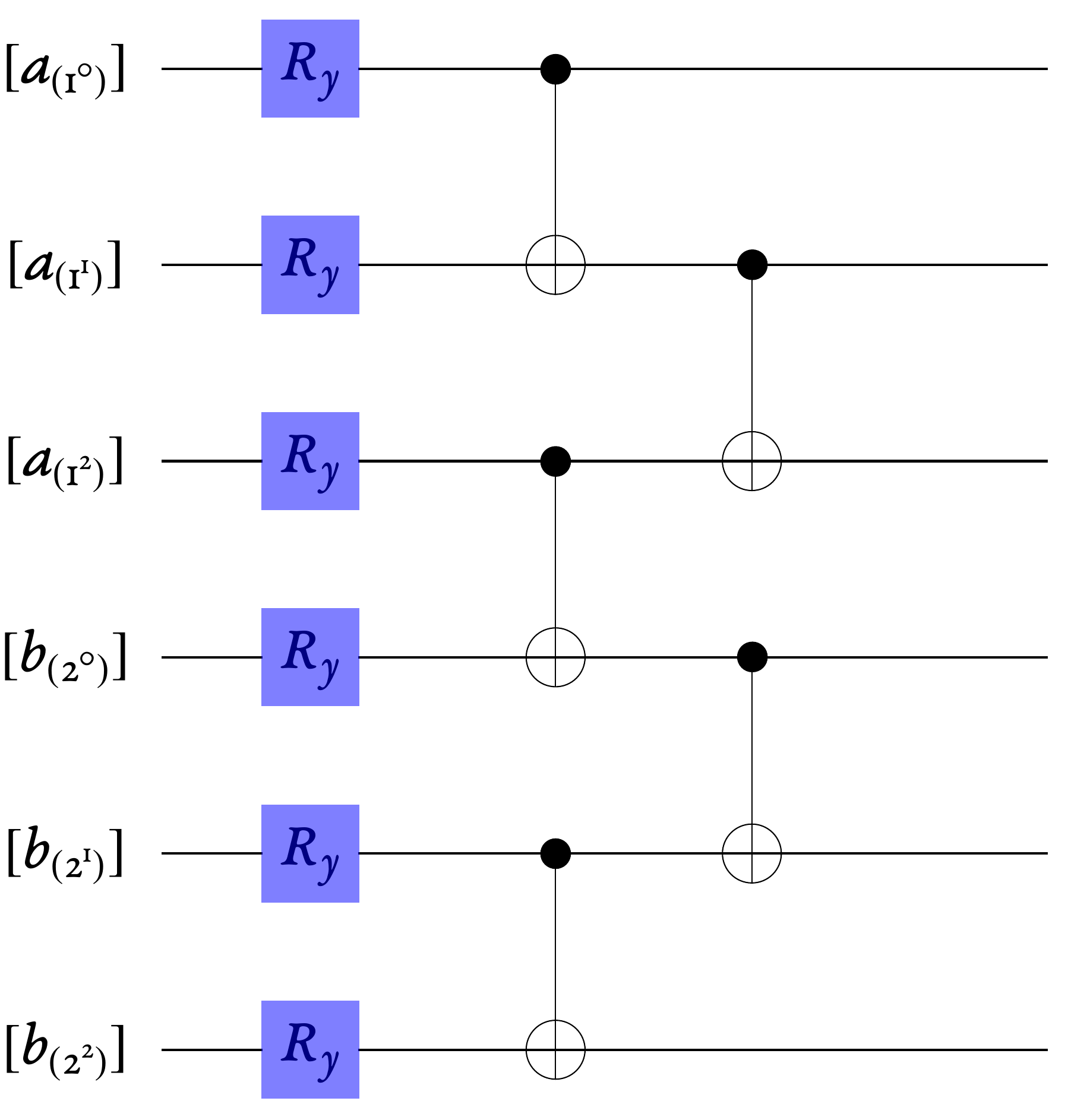}\hfill
    \includegraphics[height=0.28\textheight,width=.3\textwidth]{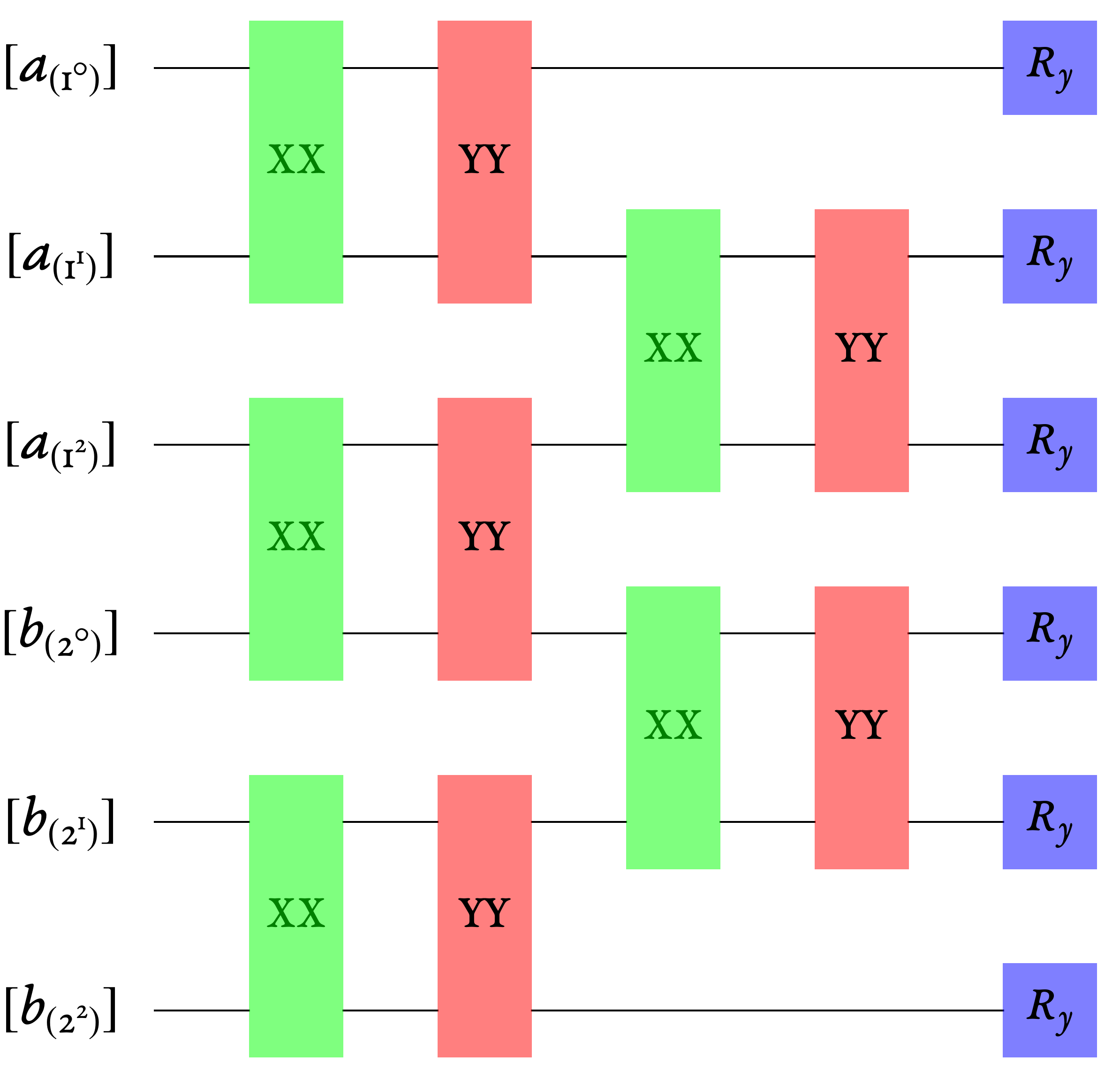}
    \caption{Representations for the circuit blocks $R_x C_z R_y$ (with nearest neighbor $C_z$ couplings only), $R_y C_x$, and $XX YY R_y$ that are used in the penalty-based VQE simulations.}
    \label{fig:penalty-circuits}
\end{figure*}

Fig. \ref{fig:penalty-circuits} shows some of the quantum circuit ansatze used in the penalty-based approach. 

% \begin{multicols}
%\lipsum[1-10]
% \end{multicols}

% \bibliographystyle{alpha}
\bibliographystyle{unsrt}
\bibliography{refs}

\end{document}